\documentstyle[12pt]{article}
\def\H4{H\hspace{-.275cm}/\hspace{.035cm}}
\def\sH4{H\hspace{-.2cm}/\hspace{.015cm}}
\def\58{\frac{15}{8}}
\def\156{\frac{15}{6}}
\def\34{\frac{3}{4}}
\def\beq{\begin{equation}}
\def\eeq{\end{equation}}
\def\bea{\begin{eqnarray}}
\def\eea{\end{eqnarray}}
\def\bq{\begin{quote}}
\def\eq{\end{quote}}
\def\ra{\rightarrow}

\begin{document}
\begin{flushright}
{IOA.327/95}\\
NTUA 53/95 \\
\end{flushright}
\begin{center}
{\bf
Gauge Coupling Unification and the Top Mass\\
     in String Models with $SU(4)\times O(4)$ Symmetry. }\\

\vspace*{1cm} {\bf G. K. Leontaris} \\
\vspace*{0.3cm}
{\it Theoretical Physics Division} \\
{\it Ioannina University} \\
{\it GR-45110 Greece} \\
\vspace*{1cm} {\bf N. D. Tracas} \\
\vspace*{0.3cm}
{\it Physics Department}\\
{\it National Technical University}\\
{\it GR-157 80 Zografou, Athens, Greece}\\
\vspace*{0.3cm}

{\bf ABSTRACT} \\
\end{center}
\noindent
We discuss the low energy implications of gauge coupling unification
at the string scale, taking into account string threshold corrections
in the
\break
$SU(4) {\times} O(4)$ model. We express $sin^2\theta_W$ and
$a_3$ as functions of the calculable string threshold differences and
discuss simple examples of spectra which retain the successful predictions
of the supersymmetric unification. Using further the low energy data
and reasonable values of the common gauge coupling at the string  scale,
we obtain the range of the threshold corrections.  Finally, we study the top
Yukawa coupling ($h_t$) evolution whose initial value is determined in
terms of the common gauge coupling at the string scale. We find that $h_t$
reaches  its (quasi) infra-red fixed point at the weak scale and discuss the
implications on the top mass.
\vspace*{.5cm}
\noindent
\begin{flushleft}
IOA-327/95 \\
NTUA 53/95\\
November 95
\end{flushleft}
\thispagestyle{empty}
\vfill\eject

\newpage

\section{Introduction}

Recent experimental evidence  indicates that the desired
unification of all fundamental forces can take place
(within a single gauge group) at a scale $M_{G}\sim 10^{16}GeV$, where
all the couplings attain a common value, provided supersymmetry
exists above a scale of the order $~ 1 TeV$. Within the context of
supersymmetry  however, the origin and magnitude of Yukawa couplings
and other parameters are not explained. Among the present candidates,
string theories can in principle give answers to the above questions.
In most of the string derived  models however, this simple unification
scenario based on a single non -- Abelian gauge group is lost.
String unification has been shown to occur at a scale some 20 times
larger than the $M_{G}$ scale predicted by the minimal supersymmetric
standard model (MSSM).
\begin{eqnarray}
M_{str}&= &g_{str}
            \frac{e^{(1-\gamma)}3^{-\frac 34}}{4\pi}M_{Pl.}
            \nonumber \\
&\approx & 5.2\,  g_{str}\times 10^{17} GeV
\end{eqnarray}
In the above, $g_{str}$ is the universal string coupling which
is fixed by the vacuum expectation value of the dilaton field
$S$, $g_{str}^2= 2/(S + \bar S)$.

The gauge symmetry of the resulting  theory below $M_{str}$ is
usually a product of groups
$G = \prod_{\alpha} G_{\alpha}$ rather than a
single  gauge group. The  corresponding
field theory describing the low energy phenomena is achieved
by integrating out the massive string states. As a result, the
evolution of the gauge couplings $g_{\alpha}$ of the effective
theory should take into account threshold corrections
$\Delta_{\alpha}$ due to the infinite tower of the massive string
modes. Thus, they are given by
\begin{eqnarray}
\frac {4\pi}{g_{\alpha}^2({\mu})}=
k_{\alpha}\frac {4\pi}{g_{str}^2} + \frac {1}{4\pi}
\left({b_{\alpha}}ln(\frac{M_{str}^2}{\mu^2}) +
 {\Delta_{\alpha}}\right)
\end{eqnarray}
where  $b_{\alpha}$ is the beta--function and
$k_{\alpha}$ characterizes the Kac-Moody level
of the corresponding coupling $g_{\alpha}$ ($k_{\alpha}=1$ in what
follows).

As it is obvious from the above formula, string thresholds
affect decisively the boundary conditions of the effective
field theory gauge couplings. Therefore, the low energy predictions
of a particular string model are also sensitive on $\Delta_{\alpha}$.

String threshold corrections have been extensively studied in
the literature \cite{ST1,ST2}. Recently there was a revived interest
from the point of view of the effective field theory\cite{ST3} and
their implications in the low energy phenomenology\cite{ST4,ST5}.

A class of string derived models\cite{aehn,alr,af,xxG,bf},
which offer a suitable ground  to study the low energy implications
of these thresholds, is based on the free fermionic formulation of
the four dimensional  superstring\cite{mb}. In the  present work, we
examine some related aspects of the string derived $SU(4)\times O(4)$
model. We explore the possibility of reconciling the low energy data
with the existence of the string unification point being twenty times
larger that the conventional unification scale.
We take into account the string threshold corrections and determine
the low energy gauge couplings in terms of their differences and
the  spectrum of the model.
We extend previous analysis  on the top mass calculations
and include the effects of the theory above the ``GUT'' scale
including the string threshold corrections.

\section{The Model}

We  briefly start  with  the basic features of the minimal
supersymmetric  version of the $SU(4)\times O(4)\sim SU(4)\times
SU(2)_L\times SU(2)_R$ model. The field content is summarized in
the following table
\begin{eqnarray}
\begin{array}{llll}
F=(4,2,1) ;&{\bar F}=({\bar 4},1,2) ;&
H=(4,1,2) ;&{\bar H}=({\bar 4},1,2) ;             \nonumber\\
h=(1,2,2) ;&D=(6,1,1);&\phi_{m,0}=(1,1,1), &m=1,2,3 ;\nonumber\\
\H4=(4,1,1) ;&{\bar \H4}=({\bar 4},1,1) ;&
a_R=(1,1,2) ;&a_L=(1,2,1) .
\label{spec}
\end{array}
\end{eqnarray}
Left  and right handed fermions (including the right handed
neutrinos) are accommodated in the  $(4,2,1)$, $ (\bar 4,1,2)$
representations  respectively.  Both pieces form up the complete
$16^{th}$  representation of  $SO(10)$.
The $SU(4)\times SU(2)_R \ra SU(3)\times U(1)$ symmetry breaking
is realized at a scale\break
$\sim 10^{15-16}GeV$, with the introduction
of a higgs pair belonging to  $H + \bar{H}= (4,1,2) + (\bar 4,1,2)$
representations. The symmetry breaking of
the standard model occurs in the presence of the two standard
doublet higgses  which are found in the $(1,2,2)$ representation of
the  original symmetry of the model. (The decomposition of the latter
under  the $SU(3)\times SU(2)_L\times U(1)_Y$ gauge group results to
the two higgs doublets $(1,2,2)\ra (1,2,\frac 12 ) + (1,2,-\frac 12)$.)
The three singlets $\phi_m$'s are engaged in the see-saw type mechanism
providing $M_G$--order masses to right handed neutrinos, while
$\phi_0$ is responsible for the appearance of the Higgs mixing term.
Finally, note the existence of the `exotic'
representations $\H4,{\bar \H4}, a_R$ and $a_L$. Although they do not
arise in the `ordinary' decomposition of an  $SO(10)$  GUT symmetry,
they do appear in string derived models constructed at the
level $k=1$ of the  Kac-Moody algebra. These states possess fractional
electric charges\cite{alr}  and are expected  to transform  non --
trivially under a hidden gauge group\cite{gl} which  becomes strong
at an intermediate scale confining them into bound states.
In our present analysis we are not going to discuss such complications.

The trilinear superpotential is
\begin{eqnarray}
{\cal W}& = & \lambda_1 F{\bar F}h   +\lambda_1^\prime F{\bar H} h
+ \lambda_2 {\bar F}H\phi_m   +\lambda_3 HHD   \nonumber     \\
&+&
\lambda_4{\bar H}{\bar H}D +
\lambda_5 FFD  +\lambda_6{\bar F}{\bar F}D+ \lambda_7 DD\phi_{m,0}
 \nonumber  \\
&+&
\lambda_8 hh\phi_{m,0}+  \lambda_9 \phi_n\phi_m\phi_l +
\lambda_{10} \phi_n\phi_m\phi_0 + \lambda_{11}\H4{\bar \H4}\phi_{m,0}
\nonumber \\
&+&
 \lambda_{12}\H4\H4 D +\lambda_{13}{\bar \H4}{\bar \H4}D +
\lambda_{14}h a_L a_R  +\lambda_{15}a_L a_L \phi_{m,0} \nonumber \\
 &+&\lambda_{16}a_R a_R \phi_{m,0}+
\lambda_{17}\H4{\bar H} a_R     +\lambda_{18}{\bar \H4} H a_R +
\lambda_{19} {\bar \H4}  F a_L \label{sp}
\end{eqnarray}
The phenomenological implications of (\ref{sp}) have been
discussed elsewhere\cite{alt,steve,ak}.
Here we will concentrate on the  renormalisation
of the gauge and Yukawa couplings  from the string scale to
low energies. From the spectrum  in (\ref{spec})
we observe first that in the minimal case   there  is
an excess of right doublet over  left doublet fields.
In fact the asymmetric form of the higgs fourplets
with respect to the two $SU(2)$ symmetries of the model, causes
a different running for the $g_{L,R}$  gauge couplings from the
string scale down to $M_G$. The possible existence of a
new pair of fourplets with $SU(2)_L$ -- transformation properties (as
suggested in ref.\cite{steve}), namely
$H_L=(4,2,1)$ and ${\bar H}_L=({\bar 4},2,1)$,
could adjust their running so as to have $g_L
= g_R$ at $M_{G}$.  This case corresponds to  family -- antifamily
representations which can become massive close to $M_{G}$, with a
trilinear or higher order term of the form $<\Phi>(\bar 4,2,1) (4,2,1)$.
Moreover, a relatively large number ($n_D$)  of sextet fields ($n_D \sim 7$)
remaining in the massless spectrum down to $M_{G}$, would also result to an
approximate equality of the above with $g_4$ coupling. Other cases of string
spectra with the desired properties are also possible.

Obviously, the equality of the three gauge couplings $g_{4,L,R}$
at the $SU(4)$ breaking scale  $M_{G}$, is of great importance.
In practice, this means that the three standard gauge couplings
$g_{1,2,3}$ start running from  $M_{G}$ down to low energies,
with the same initial condition. The only possible splitting would
arise only from string and GUT threshold corrections\cite{alt}.
Thus, choosing $M_{G}\sim 10^{16}GeV$,
we are able to obtain the correct predictions for $sin^2\theta_W$
and $a_3(m_Z)$. As a matter of fact, the intermediate gauge breaking
step gives us one more free parameter (namely $M_{G}$). Having
obtained the desired string spectrum, we are free to choose its value
in order to reconcile the high string scale $M_{str}$ with the
low energy data. Examples of  string models with such properties
have been proposed\cite{gl}.

The renormalisation group equations of the string version have
been derived and studied in previous works\cite{alt,blo,hom}.
At the one loop level, taking into account the string threshold
corrections we can obtain the following equations
\begin{eqnarray}
\frac{1}{a_3}-\frac{s^2}{\alpha}&=&
(b_4-b_L)Q_{UG}+(\vec{b}_3-\vec{b}_2){^.}\vec{Q}+
 \Delta_4-\Delta_L\\
\frac{1}{a_3}-\frac{3c^2}{5\alpha}&=&
\frac 35 (b_4-b_R)Q_{UG}+ (\vec{b}_3-\vec{b}_1){^.}\vec{Q}+
\Delta_4-\Delta_R
\end{eqnarray}
In the above, we have denoted $Q_{UG}=\frac 1{2\pi}log{(M_U/M_G)}$,
with $s,c$ the $sin$ and $cos$ of the weak mixing angle, while
 $\vec{b}_i\:^.\vec{Q}=\sum_n b_i^nQ_{n,n-1}$ takes into account
all possible intermediate scales. The weak mixing angle is given
\begin{equation}
sin^2\theta_W=\frac 38 + \frac 58\alpha\{b_{LR4}
Q_{UG}+\vec{b}_{21}\:^.\vec{Q}+\Delta_{LY}\}
\end{equation}
with
\begin{eqnarray}
b_{LR4} &=& \frac{5b_L-3b_R-2b_4}5,\quad
b_{ij} = b_i - b_j,\\
\Delta_{LY}&=& \frac{3(\Delta_L-\Delta_R)+2(\Delta_L-\Delta_4)}5
\end{eqnarray}
If we assume
the minimal supersymmetric spectrum bellow $M_G$, where only the
supersymmetry  breaking scale $M_S$ enters,  we can eliminate  $M_S$
and determine the scale $M_{G}$ in terms of $\alpha_3$, $\alpha$,
$sin^2\theta_W$ and  the differences of the string
thresholds.
Equivalently, $sin^2\theta_W$ can be expressed as follows
\begin{eqnarray}
sin^2\theta_W &=& \frac{5}{8-5\kappa}
              \{\frac 35 - \kappa\frac{\alpha}{\alpha_3}
\nonumber \\
&+&\alpha((b_{LR4}-\kappa b_{L4})Q_{UG}+(b_{21}-\kappa b_{23})Q_G
\label{s2w} \\
&-&(b_{21}^0-\kappa b_{23}^0)Q_Z + \Delta_{LY}-\kappa\Delta_{L4}
)\}
\nonumber
\end{eqnarray}
where $\Delta_{L4}=\Delta_L-\Delta_4$,
$\kappa = \frac{b_{21}^0-b_{21}}{b_{23}^0-b_{23}}$
and the superscript $0$ in the beta functions refers to the
non-supersymmetric ones.
In particular, if the beta functions $b_4,b_L,b_R$ above the GUT scale
are equal, then $b_{4L}=b_{4R} = 0$ and the above expression
for $sin^2\theta_W$ becomes very simple. In this case
the $g_{4,L,R}$ gauge coupling  splittings at $M_G$ are determined only
in terms of the differences $\Delta_{ij}$.

Before we proceed to the calculations,
 let us describe briefly the spectrum in two main energy regions.
In the $M_{str} - M_G$ region, in addition to the  three
generations of $(4,2,1)$ and $(\bar 4,1,2)$, we choose the following
content:
\begin{eqnarray}
\begin{array}{lllllllll}
 n_H = 2,& n_h =1,& n_D = 8,& n_{\sH4}=4, &n_{a_L}=4,&n_{a_R}=4,&
 n_{H_L}=2
\end{array}
\end{eqnarray}
The  above content has the property of giving $b_4=b_L=b_R=7$
(note the existence of the $H_L$'s that were mentioned before).
Therefore, in the $M_{str} - M_G$ region the three
gauge couplings $\alpha_4,\alpha_L,\alpha_R$ run in
parallel, their initial points at the $M_{str}$ scale
differing only due to the string threshold corrections
$\Delta_L,\Delta_R,\Delta_4$.
In the $M_G - M_Z$ region,
the $SU(4)\times SU(2)_L\times SU(2)_R$ model
breaks down, at $M_G$, to the MSSM, which at the scale $M_S$ turns to
the non-supersymmetric standard model SM.

The procedure we are adopting is as follows:
Using as input parameters the $M_{str}$ scale and the
string threshold differences $\Delta_{L4}=\Delta_L-\Delta_4$,
$\Delta_{R4}=\Delta_R-\Delta_4$, we determine
$M_G$ in order to have acceptable low energy parameters
$sin^2\theta_W$ and $\alpha_3$ (we fix $\alpha=1/127.9$).
As an example, we used the $\Delta_{ij}=\Delta_i-\Delta_j$ values
obtained in ref\cite{ST4} for the specific $SU(4)\times O(4)$
model\cite{alr}. In particular we take $\Delta_{L4}\sim 8.47$
and $\Delta_{R4}\sim 2.05$. We can then run the gauge couplings
from down to up and evaluate the quantities
$$\frac{1}{\alpha_{str}}+\Delta_4,\quad\quad
\frac{1}{\alpha_{str}}+\Delta_R, \quad\quad
\frac{1}{\alpha_{str}}+\Delta_L$$
The relation between  $M_{str}$ and $\alpha_{str}$ can be
used now to determine the absolute values of the string thresholds.

Let us now put our results into Figures. In Fig.1 we show the
absolute values of the three string thresholds as a function of
$sin^2\theta_W$ for $M_{str}=0.4\times 10^{18}GeV$ and $M_S=200GeV$.
We see a weak dependence on $sin^2\theta_W$.  In Fig.2 we plot the
threshold $\Delta_L$ as a function of both $sin^2\theta_W$ and
$M_{str}$. The  dependence on the latter is strong. In fact,
a change of $M_{str}$ from $0.3\times 10^{18}GeV$ to
$0.5\times 10^{18}GeV$ results
a change in $\Delta_L$ from $\sim -200$ to $\sim 100$ (in Fig.1,
we have chosen the value of $M_{str}$ which corresponds to
$\Delta_i$'s of the same order as their differences).
In other words, even large threshold corrections demand only a
small change in the value of $M_{str}$.
If $M_{str}$ is ``pushed'' towards $M_G$, then large negative
threshold corrections are required.
For completeness, in Fig.3 we plot contours of constant $M_G$ in
the plane of $(sin^2\theta_W ,a_3)$, for $M_S=200GeV$ and
$M_{str}=0.4\times 10^{18}GeV$. The line that crosses these
contours gives the acceptable pairs of $(sin^2\theta_W ,a_3)$
that correspond
to the chosen values of $M_{str}$ and $M_S$ (as we have mentioned
before we keep $\alpha=1/127.9$).

\section{Top Yukawa coupling fixed point}


In the case of the minimal supersymmetric standard model, it is well
known that the top--Yukawa coupling evolution from the unification
scale down to the low energies, exhibits a quasi--fixed point structure
\footnote{In ref.\cite{LR} it
has been shown that in  MSSM the infrared
fixed point is never reached. On the contrary, in theories with a stage
of compactification the top coupling reaches its infrared fixed
point since the evolution of couplings is much faster, following
a power low rather than a logarithmic evolution.}.
In fact, starting  at the GUT--scale with  $h_{t,G} > 1$ the top mass
is almost insensitive to the  $h_{t,G}$ value.

Large  $h_{t,G}$ values are in perfect  agreement with the recent
experimental evidence for a top mass of the ${\cal O}(170-180) GeV$.
Moreover, the idea of the $SU(2)\times U(1)$ symmetry breaking by
radiative corrections in supersymmetric theories is realized by large
negative top--Yukawa corrections to the Higgs mass, which need a
sufficiently large top coupling. The MSSM theory with the unification
assumption of the three gauge couplings at the scale $\sim 10^{16}GeV$
does not provide a convincing reason why the initial value of the
Yukawa coupling is large. It thus appears that the infra-red
structure of the top coupling has its origin in a fundamental theory
beyond the MSSM. An interesting possibility is that there is additional
structure above the supersymmetric `unification' at $M_{G}\sim
10^{16}GeV$ which determines Yukawas and other parameters at $M_{G}$.
The present string derived model provides such an example. The top
Yukawa coupling is related to the gauge coupling at the string scale
$M_{str}$. The $SU(4)$-breaking takes place at the intermediate scale
$M_{G}\sim 10^{16}GeV$, which effectively corresponds to the SUSY
`unification'
\break
scale. Knowing the evolution equations of the gauge and
Yukawas between $(M_{str}-M_{G})$, it is rather easy to determine
the $h_t$  value at the GUT scale, which will serve as initial
condition for the  $(M_{G}-M_Z)$ running.  In particular,
if the spectrum bellow the GUT SU(4) breaking scale is that of the
minimal supersymmetric standard model, then one can make a definite
prediction about the top mass.

In the present model,
all charged fermions of the third generation receive masses from the
superpotential term $\lambda_1\bar F F h$. Therefore the $SO(10)$
Yukawa unification condition $h_{(t, G)} = h_{(b, G)} = h_{(\tau ,G)}
\equiv  \lambda_1(M_G)$  is also retained in the $SU(4)\times SU(2)_L
\times  SU(2)_R$  symmetry. In the range $M_{str} - M_{G}$ the
evolution  of the Yukawa coupling $\lambda_1$ is given by
\begin{equation}
16\pi^2\frac{d\lambda_1}{dt} = \lambda_1
(8{\lambda_1}^2 - \sum_{\alpha}c_{\alpha}g_{\alpha}^2 )\label{def1}
\end{equation}
where $t=log(Q)$ is the logarithm of the scale, and the index $\alpha$
refers to the three gauge group factors $\alpha = 4,L,R$ in the range
$Q = (M_{str} - M_{G})$.  For the sake of simplicity,
in the above differential equation we have ignored terms proportional
to the other Yukawa couplings of the superpotential. If all couplings
were included, only a numerical solution would be possible\cite{wip},
however our results concerning the top-mass prediction would not be
essentially affected.
The coefficients $c_{\alpha}$ are given by
\[
\{c_{\alpha} \}_{\alpha = 4,L,R}  =
\left\{ \frac{15}2,3,{3} \right\}\nonumber
\]
Thus, in the range $M_{str}-M_{G}$ the solution for $\lambda_1$
is given by the following expression
\begin{eqnarray}
\lambda_1(t)& = &\lambda_{1}(t_{str})\gamma_U\zeta (t)\label{laA}\\
\zeta (t)&=& \frac{1}{(1 + \frac {8}{8\pi^2}
\lambda_{1}^2(t_{str})I_U(t))^{1/2}}\label{laB}
\end{eqnarray}
with
\begin{equation}
\gamma_U = \prod_{j=4,L,R}
\left(\frac{\alpha_{str}}{\alpha_{j}}
\right)^{\frac{c_j}{2b_j}}    \label{gU},\quad
I_U(t)= \int^t_{t_{str}} \gamma_U^2({t^{\prime}})
dt^{\prime} \label{IU}
\end{equation}

At the SU(4) breaking scale $M_{G}$ the original symmetry breaks
down to the standard gauge group. As pointed out previously
the top Yukawa coupling has the same initial value at $M_{G}$
with the $b-\tau$'s, i.e., we are in the case of $tan\beta\gg 1$.
In the case of the large tan$\beta$, ignoring the $\tau$ -- Yukawa,
for equal  $h_t,h_b$ couplings we can obtain the following
expression\cite{FLL} for the top-Yukawa evolution below $M_G$
\begin{eqnarray}
h_t(t)& = & \lambda_1(t_G)\gamma_Q(t)\xi (t)\label{Yta}\\
\xi (t)&=& \frac{1}{(1 + \frac {7}{8\pi^2}
 \lambda_1(t_G)^2I(t))^{1/2}}\label{Ytb}
\end{eqnarray}
where the relation $h_{t,G}\equiv \lambda_1(t_G)$ has been taken
into account.  The expressions $\gamma_Q(t), I_Q(t)$ are similar
to those of $\gamma_U, I_U$ in (\ref{IU}) respectively.
Therefore, combining the above two equations,
we determine the top Yukawa coupling and its mass at  low
energies directly from the initial value of the coupling $\lambda_1$
at $M_{str}$. In particular, imposing the initial condition
$\lambda_1(t_{str})=\sqrt{2}g_{str}$ predicted in this particular
model, we obtain the following formula for the top mass
\begin{eqnarray}
\frac{m_t(t)}{sin\beta}& = &
\sqrt{2}g_{str}\gamma_U(t_G)\zeta{(t_G)}\gamma_Q(t)\xi{(t)}
\frac{\upsilon}{\sqrt{2}}
\label{topform}
\end{eqnarray}
with $\upsilon  = 246 GeV$.

Finally, in Table I we present the (physical) top mass predictions
and the range of $tan\beta$ in order to have the running bottom mass
$m_b(m_b)=(4.15-4.35)GeV$, for three representative cases of the
supersymmetry breaking
scale $M_S$. We also show, for each case, the corresponding ``GUT''
scale $M_G$ and $\alpha_3(M_Z)$. The string scale value is
$M_{str}=0.4\times 10^{18}GeV$ ( which gives $g_{str}=0.77$) and
$sin^2\theta_W=.232$.
We have checked that the effect of string thresholds on $m_t$ is of
the order of $(4-6)\%$. Thus for given $m_b$ and $sin^2\theta_W$
(or $\alpha_3$) values, $m_t^{phys}$ is well determined in terms of
the infrared fixed property of the Yukawa coupling.

\begin{center}
$M_{str}=0.4\times 10^{18}GeV,\quad g_{str}=0.77,\quad
sin^2\theta_W=.232$\\
\vspace*{.1cm}
\[
\begin{array}{|c|c|c|c|c|}
\hline
\frac{m_t^{phys}}{GeV}  &{tan\beta}
 & \frac{M_S}{GeV} &  \frac{M_G}{GeV} & \alpha_3(M_Z)\\
\hline\hline
190         & 60-63     & 1000 &1.6\times 10^{16}& 0.122\\
187         & 58-61     & ~500 &2.0\times 10^{16}&0.124\\
183         & 56-59     & ~200 &2.5\times 10^{16}&0.127\\
\hline
\end{array}
\]
\end{center}

Note that in our actual calculations we have taken into
account the $M_S$ scale, thus running the (non supersymmetric) SM
beta-fubctions for gauge as well as Yukawa couplings. At the $M_S$
scale, the initial conditions for the $h_t$ and $h_b$ running are
of course $h_t^{NS}(t_G)=h_t(t_G)sin\beta$ and
$h_b^{NS}(t_G)=h_b(t_G)cos\beta$. We have also checked that,
running (numerically) the coupled differential
equations for $h_t$ and $h_b$ on the one hand and using the equations
(\ref{laB} --\ref{Ytb}) on the other,
the differences between these procedures are negligible. Note that
the RGEs for $h_t$ and $h_b$ in the range $M_G-M_S$ differ only in
the small U(1)-gauge coefficient.

\section{Conclusions}
In the present work we have analysed the possibility to obtain
low energy predictions compatible  with the experimental data
in string derived models with $SU(4)\times O(4)$ symmetry.
Generally, large string thresholds are required to reconcile
the experimental data with the existence of the large gap
between the string ($M_{str}\sim 5 \times 10^{17}GeV $)  and
supersymmetric ($M_{G}\sim 1.5\times 10^{16}GeV$) unification scales.
It is argued here that, a simple and viable scenario -- compatible with
the low energy phenomenological expectations -- is to obtain a massless
spectrum which  allows approximatelly a  parallel evolution of the
gauge couplings  between $M_{str}  -  M_{G}$.
Given the rich spectrum of such models\cite{aehn,alr,af}, one could
choose carefully the vacuum expectation values of the singlet fields
associated with the large breaking scale of the possible surplus $U(1)$
symmetries and make massive those states which allow equal beta
function coefficients in most of the range  above $ M_G$.
This would simply correspond to a judicious choice of a specific
flat direction of the effective field theory superpotential.

In the above context, we have considered the evolution of the top
Yukawa coupling from the string scale down to low energies.
We have found that it exhibits a fixed point structure thus
leading to definite predictions for the top mass compatible
with its present experimentally determined range.

\newpage
\begin{center}
{\large {\bf Figure Captions}}
\end{center}

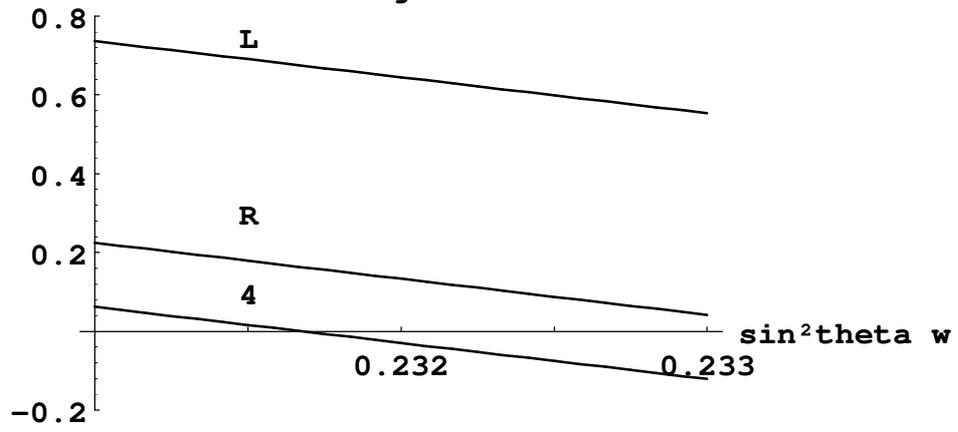
\noindent {\bf Fig.1}\\
The $\Delta_L$, $\Delta_R$ and $\Delta_4$ thresholds as a function of
$sin^2\theta_W$, for $M_{str}=0.4\times 10^{18}GeV$ and $M_S=200GeV$.

\noindent{\bf Fig.2}\\
The $\Delta_L$ threshold as a function of $sin^2\theta_W$ and
$M_{str}$. The supersymmetry breaking scale is $M_S=200GeV$.

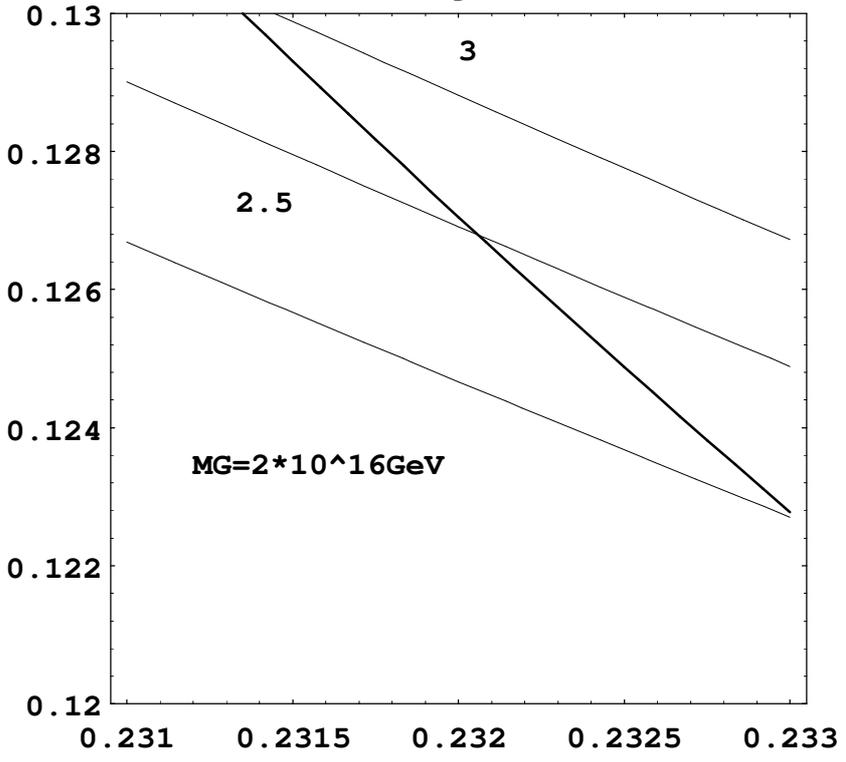
\noindent{\bf Fig.3}\\
Contours of constant $M_G=(2, 2.5, 3)\times 10^{16}GeV$ in the
$(sin^2\theta_W,\alpha_3(M_Z))$ plane. The line crossing these
contours gives the acceptable $(sin^2\theta_W,\alpha_3(M_Z))$ pairs which
correspond to the chosen values of $M_{str}=0.4\times 10^{18}GeV$
and $M_S=200GeV$ ($\alpha=1/127.9$).

\newpage


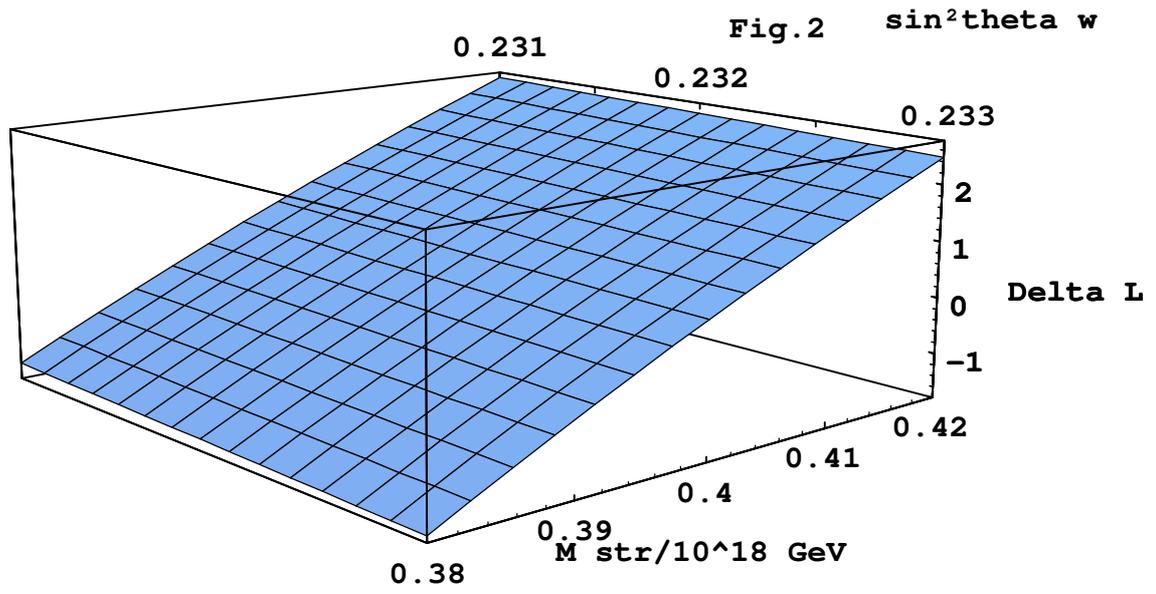
\end{document}